\documentclass[11pt]{article}
\usepackage{graphicx}
\usepackage{epsfig}
\usepackage{amsmath,amssymb}

\newcommand{\cn}{\mathrm{cn}}

\newcommand{\sech}{\rm sech}
\newcommand{\csch}{\rm csch}

\usepackage[labelsep=space]{caption}
\begin{document}
\title{Weierstrass type projective Riccati equation
expansion method and solutions of KdV equation}

\author{Sirendaoreji\footnote{E-mail: siren@imnu.edu.cn}\\{\small
       Mathematical Science College, Inner Mongolia Normal University,}\\
       {\small Huhhot 010022, Inner Mongolia, P. R. China}}
\date{}
\maketitle
\begin{abstract}
This paper presents two new Weierstrass elliptic function solutions
of the projective Riccati equations and four conversion formulas for
converting the Weierstrass elliptic functions to the hyperbolic
and trigonometric functions. The Weierstrass elliptic function
solutions to the projective Riccati equations and the conversion
formulas are used to propose the called Weierstrass type projective
Riccati equation expansion method.
The Weierstrass elliptic function solutions, the solitary wave and
the periodic wave solutions of the KdV equation are constructed by
using the proposed method.
The solitary wave like and the periodic wave solutions of the KdV
equation are shown through some figures.
\end{abstract}

\section{Introduction}
\label{intro}
\sloppy
Exact solutions to nonlinear evolution equations (NLEEs) play an important role
in applied mathematics, mathematical physics and nonlinear science. Therefore,
the problem of how to find new methods for solving NLEEs has become a hot topic
in soliton theory. In the past decades, various direct methods such as the
tanh--function method~\cite{RefJ-1}, the auxiliary equation method~\cite{RefJ-2},
the Riccati equation expansion method~\cite{RefJ-3}, the unified Riccati equation
expansion method~\cite{RefJ-4}, the Jacobi elliptic function expansion
method~\cite{RefJ-5}, the Weierstrass elliptic function method~\cite{RefJ-6a,RefJ-6b,RefJ-6c,RefJ-6d,RefJ-6e} and the projective Riccati
equation expansion method~\cite{RefJ-7a,RefJ-7b,RefJ-7c,RefJ-7d,RefJ-7e}, and so on,
have been proposed to find exact traveling wave solutions of NLEEs. Among them,
the projective Riccati equation expansion method can usually give some new special
solutions of NLEEs. Therefore, the researchers give more attention to find new
solutions of the projective Riccati equation. So far, two sets of Weierstrass
elliptic function solutions of the projective Riccati equation have been given and
applied to construct the Weierstrass elliptic function solutions of NLEEs~\cite{RefJ-8a,RefJ-8b,RefJ-8c}. But most of these Weierstrass elliptic
function solutions cannot be transformed into the hyperbolic and trigonometric
function solutions because the previously used conversion formulas were
determined by the roots of the third order polynomial equation $w^3-g_2w-g_3=0$.
Therefore, this paper aims to give more Weierstrass elliptic function solutions of
the projective Riccati equations and establish the so--called conversion formulas
which can transform the Weierstrass elliptic function solutions of NLEEs into the
hyperbolic and trigonometric function solutions. Further, we shall use the Weierstrass
elliptic function solutions of the projective Riccati equations and the conversion
formulas to propose a direct method which we named as the Weierstrass type projective
Riccati equation expansion method. Finally, we shall take the KdV equation as an
illustrative example to show the effectiveness of our method.\par
This paper is organized as follows. In the next section, four sets of Weierstrass
elliptic function solutions of the projective Riccati equations and four conversion
formulas to transform the Weierstrass elliptic function into the hyperbolic and
trigonometric function are constructed. In addition, the Weierstrass type projective
Riccati equation expansion method is proposed to find exact traveling wave solutions
to NLEEs.
In Sec.\ref{sec:3}, the proposed method is applied to construct the Weierstrass
elliptic function solutions, the solitary and periodic wave solutions of the KdV
equation. The conclusions are given in Sec.\ref{sec:4}.
\section{Weierstrass type projective Riccati equation expansion method}
\label{sec:2}
The Weierstrass elliptic function $w=\wp(\xi,g_2,g_3)$ is defined as the inverse
function of the Weierstrass elliptic integral
\begin{align}
\label{eq2:1}
\xi=\int_\infty^w {\frac {dt}{\sqrt{4t^3-g_2t-g_3}}},
\end{align}
or the solution of the following nonlinear ordinary differential equation
\begin{align}
\label{eq2:2}
{\frac {dw}{d\xi}}=4w^3-g_2w-g_3,
\end{align}
where $g_2,g_3$ are real parameters and called invariants.\par
In the following we shall consider the projective Riccati equations of the form
\begin{align}
\label{eq2:3}
\left\{\begin{aligned}
&F^{\prime}(\xi)=pF(\xi)G(\xi),\\
&G^{\prime}(\xi)=q+pG^2(\xi)-rF(\xi),
\end{aligned}\right.
\end{align}
where $F$,$G$ are unknown functions of the variable $\xi$ and $p,q,r$
are constants. By using the direct assumption approach we can construct
the Weierstrass elliptic function solutions of Eqs.(\ref{eq2:3}) as following
\begin{align}
\label{eq2:4}
&\left\{\begin{aligned}
   &F(\xi)={\frac q{6r}}+{\frac 2{pr}}\wp(\xi,g_2,g_3),\\
   &G(\xi)={\frac {12\wp^\prime(\xi,g_2,g_3)}{p\left[pq+12\wp(\xi,g_2,g_3)\right]}},
   \end{aligned}\right.\\
   \label{eq2:5}
&{\qquad}G^2(\xi)=-{\frac qp}+{\frac {2r}{p}}F(\xi),\\
\label{eq2:6}
&\left\{\begin{aligned}
&F(\xi)={\frac {5q}{6r}}+{\frac {5q^2p}{72r\wp(\xi,g_2,g_3)}},\\
&G(\xi)={\frac {-q\wp^\prime(\xi,g_2,g_3)}{\left[qp+12\wp(\xi,g_2,g_3)\right]\wp(\xi,g_2,g_3)}},
\end{aligned}\right.\\
\label{eq2:7}
&{\qquad}G^2(\xi)=-{\frac qp}+{\frac {2r}{p}}F(\xi)-{\frac {24r^2}{25pq}}F^2(\xi),
\end{align}
\begin{align}
\label{eq2:8}
&\left\{\begin{aligned}
&F(\xi)={\frac {(2+q)\left[pq+12\wp(\xi,g_2,g_3)\right]}{r\left[12p+pq+12\wp(\xi,g_2,g_3)\right]}},\\
&G(\xi)={\frac {\wp^\prime(\xi,g_2,g_3)}{\left(\wp(\xi,g_2,g_3)+{\frac p2}
+{\frac {pq}{12}}\right)^2-{\frac {p^2}{4}}}},
\end{aligned}\right.\\
\label{eq2:9}
&{\qquad}G^2(\xi)=-{\frac qp}+{\frac {2r}{p}}F(\xi)-{\frac {r^2(q+4)}{p(q+2)^2}}F^2(\xi),\\
\label{eq2:10}
&\left\{\begin{aligned}
&F(\xi)={\frac {q(p^2+2)\left[pq+12\wp(\xi,g_2,g_3)\right]}{pr\left[12q+p^2q+12p\wp(\xi,g_2,g_3)\right]}},\\
&G(\xi)={\frac {q\wp^\prime(\xi,g_2,g_3)}{\left(p\wp(\xi,g_2,g_3)+{\frac q2}
+{\frac {p^2q}{12}}\right)^2-{\frac {q^2}{4}}}},
\end{aligned}\right.\\
\label{eq2:11}
&{\qquad}G^2(\xi)=-{\frac qp}+{\frac {2r}{p}}F(\xi)-{\frac {pr^2(p^2+4)}{q(p+2)^2}}F^2(\xi),
\end{align}
where the invariants $g_2$ and $g_3$ are given by
\begin{align}
\label{eq2:12}
g_2={\frac {p^2q^2}{12}},g_3={\frac {p^3q^3}{216}}.
\end{align}
Here we point out that the solutions (\ref{eq2:4}) and (\ref{eq2:6}) are
previously known~\cite{RefJ-8a,RefJ-8b,RefJ-8c}, but the solutions (\ref{eq2:8}) and
(\ref{eq2:10}) are new.\par
We find that the Weierstrass elliptic function can degenerate to the hyperbolic and
trigonometric function by using the following conversion formulas
\begin{align}
\label{eq2:13}
&\wp(\xi,{\frac {\theta^2}{12}},-{\frac {\theta^3}{216}})
={\frac {\theta}{12}}-{\frac {\theta}{4}}\sech^2({\frac {\sqrt{\theta}}{2}}\,\xi),\theta>0,\\
\label{eq2:14}
&\wp(\xi,{\frac {\theta^2}{12}},-{\frac {\theta^3}{216}})
={\frac {\theta}{12}}+{\frac {\theta}{4}}\csch^2({\frac {\sqrt{\theta}}{2}}\,\xi),\theta>0,\\
\label{eq2:15}
&\wp(\xi,{\frac {\theta^2}{12}},-{\frac {\theta^3}{216}})
={\frac {\theta}{12}}-{\frac {\theta}{4}}\sec^2({\frac {\sqrt{-\theta}}{2}}\,\xi),\theta<0,\\
\label{eq2:16}
&\wp(\xi,{\frac {\theta^2}{12}},-{\frac {\theta^3}{216}})
={\frac {\theta}{12}}-{\frac {\theta}{4}}\csc^2({\frac {\sqrt{-\theta}}{2}}\,\xi),\theta<0,
\end{align}
where $\theta$ is a real number.\par
A given NLEE with respect to two variables $x$ and $t$ is of the form
\begin{align}
\label{eq2:17}
P(u,u_x,u_t,u_{xx},u_{xt},u_{tt},u_{xxx},\cdots)=0,
\end{align}
where the subscripts denote the partial derivatives,$P$ is a polynomial in unknown
function $u(x,t)$ and its derivatives.\par
The Weierstrass type projective Riccati equation expansion method proposed
here follows the following five steps.\par
{\em Step 1:} Making the wave transformation
\begin{align}
\label{eq2:18}
u(x,t)=u(\xi),\xi=x-{\omega}t,
\end{align}
we may exchange the Eq. (\ref{eq2:17}) into the following ODE
\begin{align}
\label{eq2:19}
H(u,u^\prime,u^{\prime\prime},\cdots)=0,
\end{align}
where primes denote the derivatives with respect to $\xi$.\par
{\em Step 2:} Assume that the Eq. (\ref{eq2:19}) has the truncated formal
series solution
\begin{align}
\label{eq2:20}
u(\xi)=a_0+\sum_{i=1}^n F^{i-1}(\xi)\left(a_iF(\xi)+b_iG(\xi)\right),
\end{align}
where $F(\xi)$ and $G(\xi)$ are the Weierstrass elliptic function solution
of the projective Riccati Eqs. (\ref{eq2:3}), $n$ is an integer number which
can be determined by balancing the highest order derivative terms with the
highest power nonlinear terms in (\ref{eq2:19}),$a_0$,$a_i,b_i~(i=1,2,\cdots,n)$
are undetermined constants and $a_n^2+b_n^2\not=0$.
\par
{\em Step 3:} Substituting (\ref{eq2:20}), (\ref{eq2:3}) together with one
of the relations (\ref{eq2:5}), (\ref{eq2:7}), (\ref{eq2:9}), (\ref{eq2:11})
into (\ref{eq2:19}) and equating the coefficients of like
powers of $F^i(\xi)G^j(\xi)$ to zero yields a set of algebraic equations.
Solving this set of algebraic equations with the aid of Maple or Mathematica
we can determine the values of $\omega,a_0$ and $a_i,b_i~(i=1,2,\cdots,n)$.\par
{\em Step 4:} Putting each solutions of the algebraic equations obtained in
{\em Step 3} together with the Weierstrass elliptic function solutions $F(\xi),G(\xi)$
of the projective Riccati equations into (\ref{eq2:20}) and using (\ref{eq2:18}),
we obtain the Weierstrass type traveling wave solutions of the Eq. (\ref{eq2:17}).\par
{\em Step 5:} Taking the conversion formulas into the Weierstrass elliptic function
type traveling wave solutions obtained in {\em Step 4}, we get the solitary wave
and the periodic wave solutions of the Eq. (\ref{eq2:17}).

\section{Solutions of the KdV equation}
\label{sec:3}
Now let us consider the famous KdV equation
\begin{align}
\label{eq3:1}
u_t+6uu_x+u_{xxx}=0.
\end{align}
Taking the wave transformation (\ref{eq2:18}) into (\ref{eq3:1})
we get the following ODE
\begin{align}
\label{eq3:2}
-{\omega}u^\prime(\xi)+6u(\xi)u^\prime(\xi)+u^{\prime\prime\prime}(\xi)=0.
\end{align}
By using the homogenous balance method we can determine that $n=2$.Thus
the solution of Eq. (\ref{eq3:2}) can be chosen as
\begin{align}
\label{eq3:3}
u(\xi)=a_0+a_1F(\xi)+b_1G(\xi)+a_2F^2(\xi)+b_2F(\xi)G(\xi),
\end{align}
where $F(\xi),G(\xi)$ are the solution of the projective Riccati equations
(\ref{eq2:3}),$a_0,a_1,a_2,b_1,b_2$ are undermined constants.\par
(1)\;Taking (\ref{eq3:3}) with (\ref{eq2:3}),(\ref{eq2:5}) into (\ref{eq3:2})
and setting the coefficients of $F^iG^j~(i=1,2,3,4;j=0,1)$ to zero we
obtain a set of algebraic equations
\begin{align*}
\left\{\begin{aligned}
&12a_2^2p=0,42a_2b_2r=0, \\
&30a_2p^2r+18a_1a_2p+18b_2^2r=0,\\
&30b_2pr^2+30a_1b_2r+30a_2b_1r-18a_2b_2q=0,\\
&-a_1p^2q+6a_0a_1p-{\omega}a_1p+6b_1^2r-6b_1b_2q=0,\\
&-b_1pqr+b_2pq^2+6a_0b_1r-6a_0b_2q-6a_1b_1q-{\omega}b_1r+{\omega}b_2q=0,\\
&3b_1pr^2-15b_2pqr+18a_0b_2r+18a_1b_1r-12a_1b_2q-12a_2b_1q-3{\omega}b_2r=0,\\
&6a_1p^2r-8a_2p^2q+12a_0a_2p+6a_1^2p-2{\omega}a_2p+24b_1b_2r-6b_2^2q=0.
\end{aligned}\right.
\end{align*}
Solving this set of algebraic equations with use of Maple we get
\begin{align}
\label{eq3:4}
a_0={\frac 16}\left(qp+\omega\right),a_1=-pr,a_2=0,b_1=0,b_2=0.
\end{align}
Substituting (\ref{eq3:4}),(\ref{eq2:4}),(\ref{eq2:20}) into (\ref{eq3:3}) we
obtain the Weierstrass type traveling wave solution of the KdV equation
\begin{align}
\label{eq3:5}
u(x,t)={\frac {\omega}{6}}-2\wp(x-{\omega}t,g_2,g_3),
\end{align}
where the invariants $g_2,g_3$ are determined by (\ref{eq2:12}).\par
To take the conversion formula (\ref{eq2:13})--(\ref{eq2:16}) with $\theta=-qp$
into (\ref{eq3:5}) we obtain the solitary wave and periodic wave solutions of
the KdV equation
\begin{align*}
&u_1^{(1)}(x,t)={\frac {qp+\omega}{6}}-{\frac {qp}{2}}\,\sech^2{\frac {\sqrt{-qp}}{2}}(x-{\omega}t),qp<0,\\
&u_2^{(1)}(x,t)={\frac {qp+\omega}{6}}+{\frac {qp}{2}}\,\csch^2{\frac {\sqrt{-qp}}{2}}(x-{\omega}t),qp<0,\\
&u_3^{(1)}(x,t)={\frac {qp+\omega}{6}}-{\frac {qp}{2}}\,\sec^2{\frac {\sqrt{qp}}{2}}(x-{\omega}t),qp>0,\\
&u_4^{(1)}(x,t)={\frac {qp+\omega}{6}}-{\frac {qp}{2}}\,\csc^2{\frac {\sqrt{qp}}{2}}(x-{\omega}t),qp>0,
\end{align*}
where $p,q,\omega$ are constants.\par
(2)\;Taking (\ref{eq3:3}) with (\ref{eq2:3}), (\ref{eq2:7}) into (\ref{eq3:2})
and setting the coefficients of $F^iG^j~(i=1,2,3,4,5;j=0,1)$ to zero
we obtain the following set of algebraic equations
\begin{align*}
\left\{\begin{aligned}
&-{\frac {576a_2b_2r^2}{25q}}+{\frac {13824b_2pr^4}{625q^2}}=0,\\ &-{\frac {288b_2^2r^2}{25q}}-{\frac {576a_2p^2r^2}{25q}}+12a_2^2p=0,\\
&18a_2a_1p-{\frac {144a_1p^2r^2}{25q}}-{\frac {432b_1b_2r^2}{25q}}+18b_2^2r+30a_2p^2r=0,\\
&-{\frac {432b_2a_1r^2}{25q}}-{\frac {432b_1a_2r^2}{25q}}
   +{\frac {3456b_1pr^4}{625q^2}}-{\frac {288b_2pr^3}{5q}}+42a_2b_2r=0,\\
&-2{\omega}a_2p+12a_0a_2p+24b_1b_2r-{\frac {144b_1^2r^2}{25q}}
   +6a_1^2p-6b_2^2q+6a_1p^2r-8a_2p^2q=0,\\
&{\frac {48{\omega}b_2r^2}{25q}}-{\frac {288b_1a_1r^2}{25q}}
   -{\frac {288a_0b_2r^2}{25q}}-{\frac {288b_1pr^3}{25q}}
   +{\frac {246b_2pr^2}{5}}-18a_2b_2q\\
&{\quad}+30a_2b_1r+30a_1b_2r=0,\\
&{\frac {24{\omega}b_1r^2}{25q}}-{\frac {144a_0b_1r^2}{25q}}
   +{\frac {171b_1pr^2}{25}}-15b_2rpq-3{\omega}b_2r-12a_2b_1q
   -12a_1b_2q\\
&{\quad}+18a_1b_1r+18a_0b_2r=0,\\
&-a_1p^2q+6a_0a_1p-{\omega}a_1p+6b_1^2r-6b_1b_2q=0,\\
&-b_1pqr+b_2pq^2+6a_0b_1r-6a_0b_2q-6a_1b_1q-{\omega}b_1r+{\omega}b_2q=0.
\end{aligned}\right.
\end{align*}
Using Maple to solve above algebraic equations we obtain
\begin{align}
\label{eq3:6}
a_0={\frac {pq+\omega}{6}},a_1=-pr,a_2={\frac {24pr^2}{25q}},b_1=0,
b_2=2\varepsilon{pr}\sqrt{-\frac {6p}{25q}},
\end{align}
where $\varepsilon=\pm{1}$ and $pq<0$.\par
Substituting (\ref{eq3:6}), (\ref{eq2:6}) and (\ref{eq2:18}) into (\ref{eq3:3})
leads the Weierstrass type traveling wave solution of the KdV equation
\begin{align}
\label{eq3:7}
u(x,t)={\frac {9\left[4\omega\wp(\xi)+p^2q^2\right]\wp(\xi)+pq\left[p^2q^2
    -6\varepsilon\sqrt{-6pq}\wp^\prime(\xi)\right]}{216\wp^2(\xi)}},
\end{align}
where $pq<0$,$\wp(\xi)=\wp(x-{\omega}t,g_2,g_3)$, the invariants $g_2$,$g_3$
are determined by (\ref{eq2:12}).\par
Because $pq<0$, so it is known from the conversion formulas that the
Weierstrass function solution (\ref{eq3:7}) cannot degenerate to the
periodic solutions. Thus, substituting the conversion formulas (\ref{eq2:13})
and (\ref{eq2:14}) with $\theta=-qp$ into (\ref{eq3:7}) leads the solitary
wave like solutions of the KdV equation as following
\begin{align*}
&u_1^{(2)}(x,t)={\frac {(pq+\omega)\cosh^4\eta+3(3pq-2\omega)\cosh^2\eta
    -6\sqrt{6}{\varepsilon}pq\sinh\eta\cosh\eta+9\omega}
    {6\left(\cosh^2\eta-3\right)^2}},\\
&u_2^{(2)}(x,t)={\frac {(pq+\omega)\sinh^4\eta-3(3pq-2\omega)\sinh^2\eta
    +6\sqrt{6}{\varepsilon}pq\sinh\eta\cosh\eta+9\omega}
    {6\left(\sinh^2\eta+3\right)^2}},
\end{align*}
where $\eta={\frac 12}\sqrt{-pq}(x-{\omega}t)$,$pq<0$.\par
(3)\; Inserting (\ref{eq3:3}) with (\ref{eq2:3}), (\ref{eq2:9}) into (\ref{eq3:2})
and setting the coefficients of $F^iG^j~(i=1,2,3,4,5;j=0,1)$ to zero
we obtain the following set of algebraic equations
\begin{align*}
\left\{\begin{aligned}
&12a_2^2p-{\frac {48b_2^2r^2+96a_2p^2r^2+12b_2^2r^2q+24a_2p^2r^2q}{(q+2)^2}}=0,\\
&{\frac {384b_2pr^4+24b_2pr^4q^2+192b_2pr^4q}{(q+2)^4}}
   -{\frac {96a_2b_2r^2+24a_2b_2r^2q}{(q+2)^2}}=0,\\
&18a_2a_1p-{\frac {24a_1p^2r^2+72b_1b_2r^2+6a_1p^2r^2q+18b_1b_2r^2q}{(q+2)^2}}
   +18b_2^2r+30a_2p^2r=0,\\
&12a_0a_2p-2{\omega}a_2p+24b_1b_2r+6a_1^2p-6b_2^2q+6a_1p^2r-8a_2p^2q
   -{\frac {6b_1^2r^2q+24b_1^2r^2}{(q+2)^2}}=0,\\
&42a_2b_2r-{\frac {18b_2a_1r^2q+18b_1a_2r^2q+65b_2pr^3q+260b_2pr^3+72b_2a_1r^2+72b_1a_2r^2}{(q+2)^2}}\\
&{\quad}+{\frac {6b_1pr^4q^2+48b_1pr^4q+96b_1pr^4}{(q+2)^4}}=0,\\
&{\frac {4b_1pr^2q^2+16b_1pr^2q+4{\omega}b_1r^2-24a_0b_1r^2+{\omega}b_1r^2q-6a_0b_1r^2q}{(q+2)^2}}-19b_2prq+6b_1pr^2\\
&{\quad}-3{\omega}b_2r-12a_2b_1q-12a_1b_2q+18a_1b_1r+18a_0b_2r=0,\\
&{\frac {2{\omega}b_2r^2q-14b_1pr^3q-12b_1a_1r^2q-12a_0b_2r^2q-56b_1pr^3}{(q+2)^2}}-18a_2b_2q+30a_2b_1r\\
&{\quad}+30a_1b_2r+39b_2pr^2
+{\frac {20b_2pr^2q^2+80b_2pr^2q+8{\omega}b_2r^2-48b_1a_1r^2-48a_0b_2r^2}{(q+2)^2}}=0,\\
&-a_1p^2q+6a_0a_1p-{\omega}a_1p+6b_1^2r-6b_1b_2q=0,\\
&-2b_1pqr+b_2pq^2+6a_0b_1r-6a_0b_2q-6a_1b_1q-{\omega}b_1r+{\omega}b_2q=0.
\end{aligned}\right.
\end{align*}
Solving this set of algebraic equations with aid of Maple we obtain
\begin{align}
\label{eq3:8}
a_0={\frac {\omega}{6}}-{\frac {2p}{3}}, a_1=-pr, a_2=0,b_1=0, b_2=0, q=-4.
\end{align}
From which we calculate that
\begin{align}
\label{eq3:9}
g_2={\frac {4p^2}{3}},g_3=-{\frac {8p^3}{27}}.
\end{align}
Substituting (\ref{eq3:8}) with (\ref{eq3:9}),(\ref{eq2:8}) into (\ref{eq3:3})
we get the Weierstrass type traveling wave solution of the KdV equation
\begin{align}
\label{eq3:10}
u(x,t)={\frac {\omega}{6}}-{\frac {2p}{3}}
  -{\frac {2p\left[p-3\wp(x-{\omega}t,g_2,g_3)\right]}{2p+3\wp(x-{\omega}t,g_2,g_3)}},
\end{align}
where the invariants $g_2,g_3$ are given by (\ref{eq2:12}).\par
Taking the conversion formulas (\ref{eq2:13})--(\ref{eq2:16}) with
$\theta=4p$ into (\ref{eq3:10}) leads the solitary and periodic wave solutions
of the KdV equation
\begin{align*}
&u_1^{(3)}(x,t)={\frac {\omega-4p}{6}}-2p\,\csch^2\sqrt{p}(x-{\omega}t),p>0,\\
&u_2^{(3)}(x,t)={\frac {\omega-4p}{6}}+2p\,\sech^2\sqrt{p}(x-{\omega}t),p>0,\\
&u_3^{(3)}(x,t)={\frac {\omega-4p}{6}}+2p\,\csc^2\sqrt{-p}(x-{\omega}t),p<0,\\
&u_4^{(3)}(x,t)={\frac {\omega-4p}{6}}+2p\,\sec^2\sqrt{-p}(x-{\omega}t),p<0.
\end{align*}
We can see that the above solutions $u_i^{(3)}\,(1,2,3,4)$ are the special
case $q=-4$ of the solutions $u_i^{(1)}\,(i=2,1,4,3)$. Here, we need to pointed
out that the case of other NLEEs, the Weierstrass elliptic function solutions
(\ref{eq2:4}) and (\ref{eq2:8}) usually can give different types of solutions.
Therefore, the solutions (4) and (8) are the different solutions of the projective
Riccati equation.
\begin{figure}[h]
\centering
\includegraphics[width=0.30\textwidth]{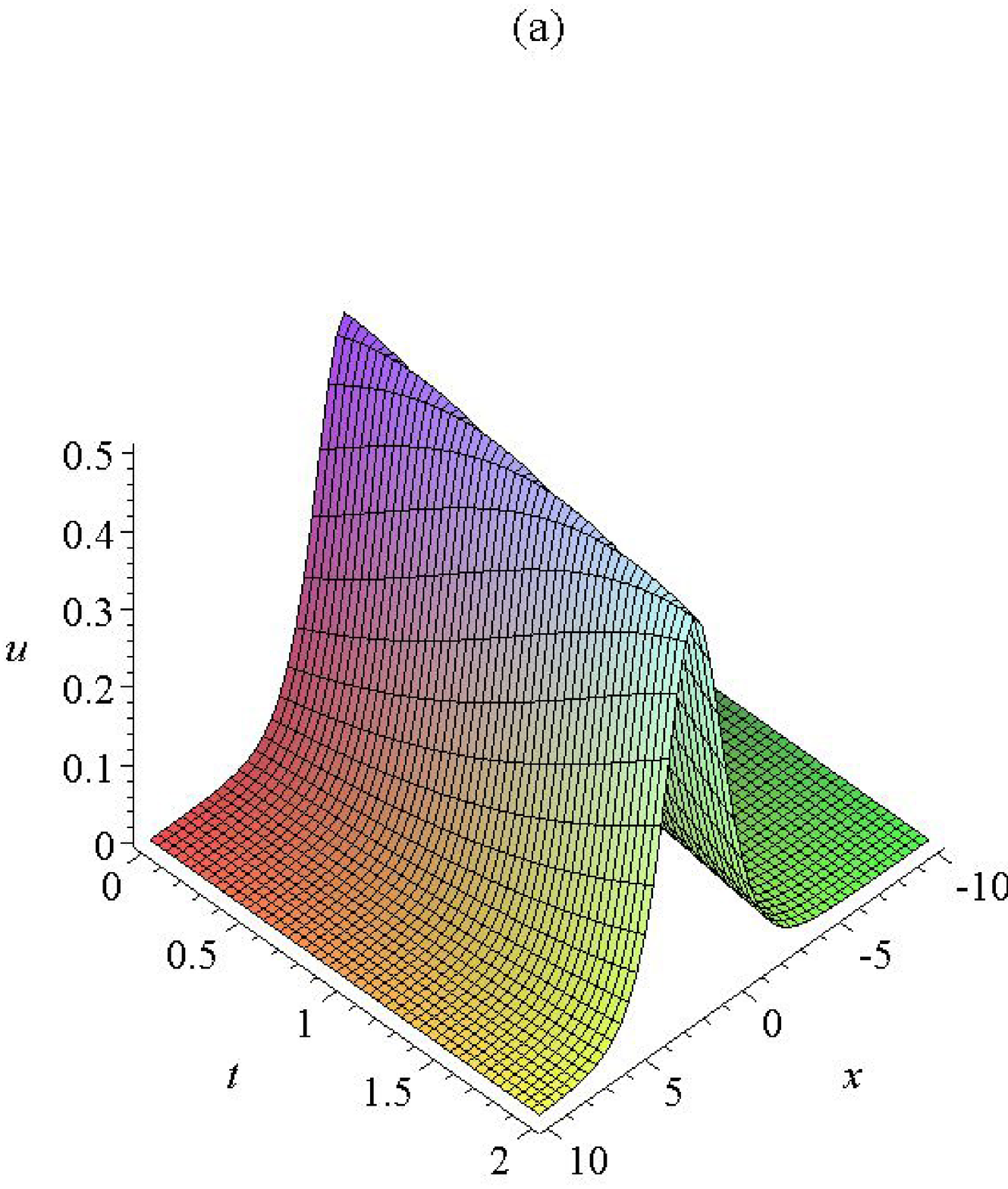}
\includegraphics[width=0.30\textwidth]{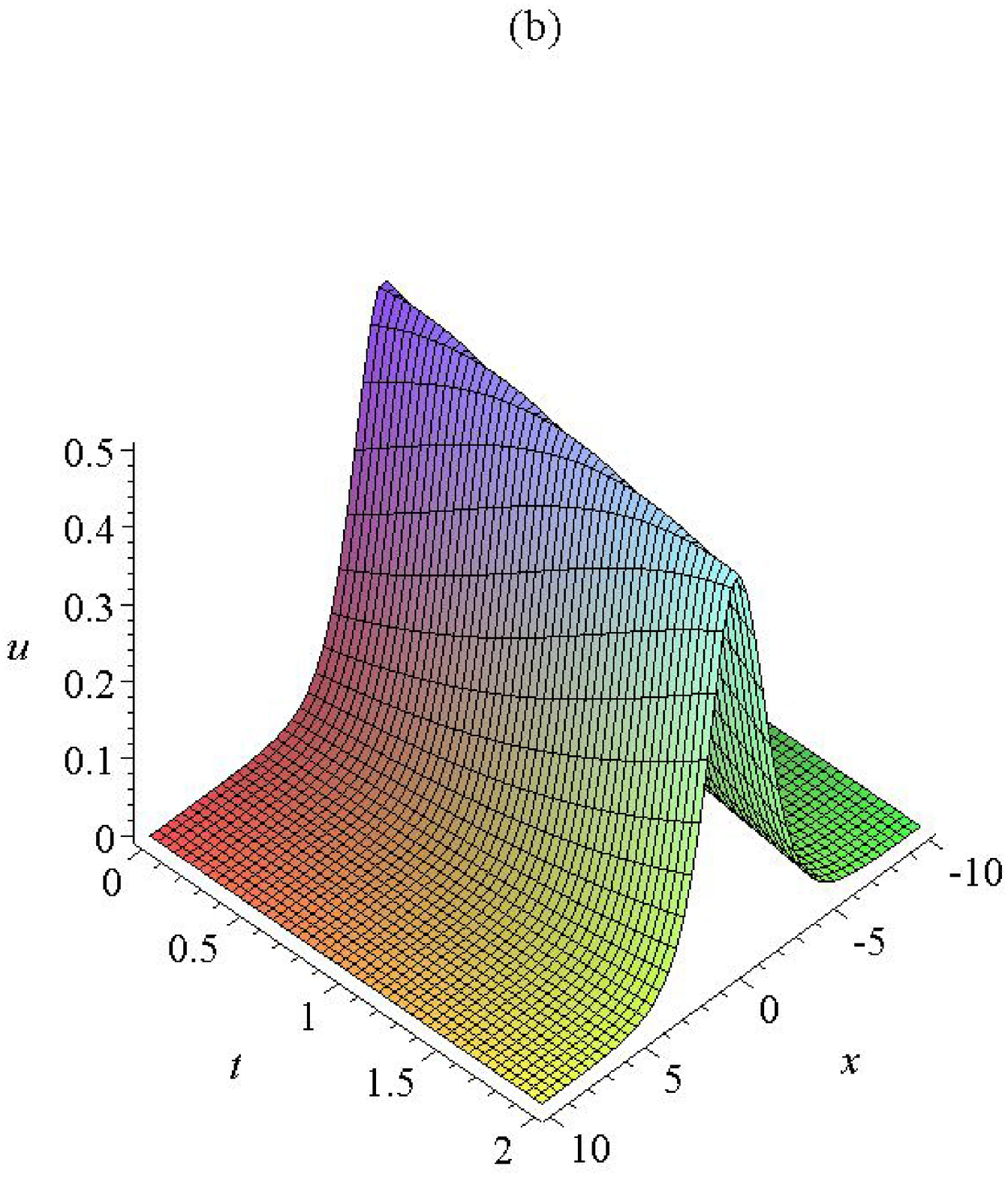}
\includegraphics[width=0.30\textwidth]{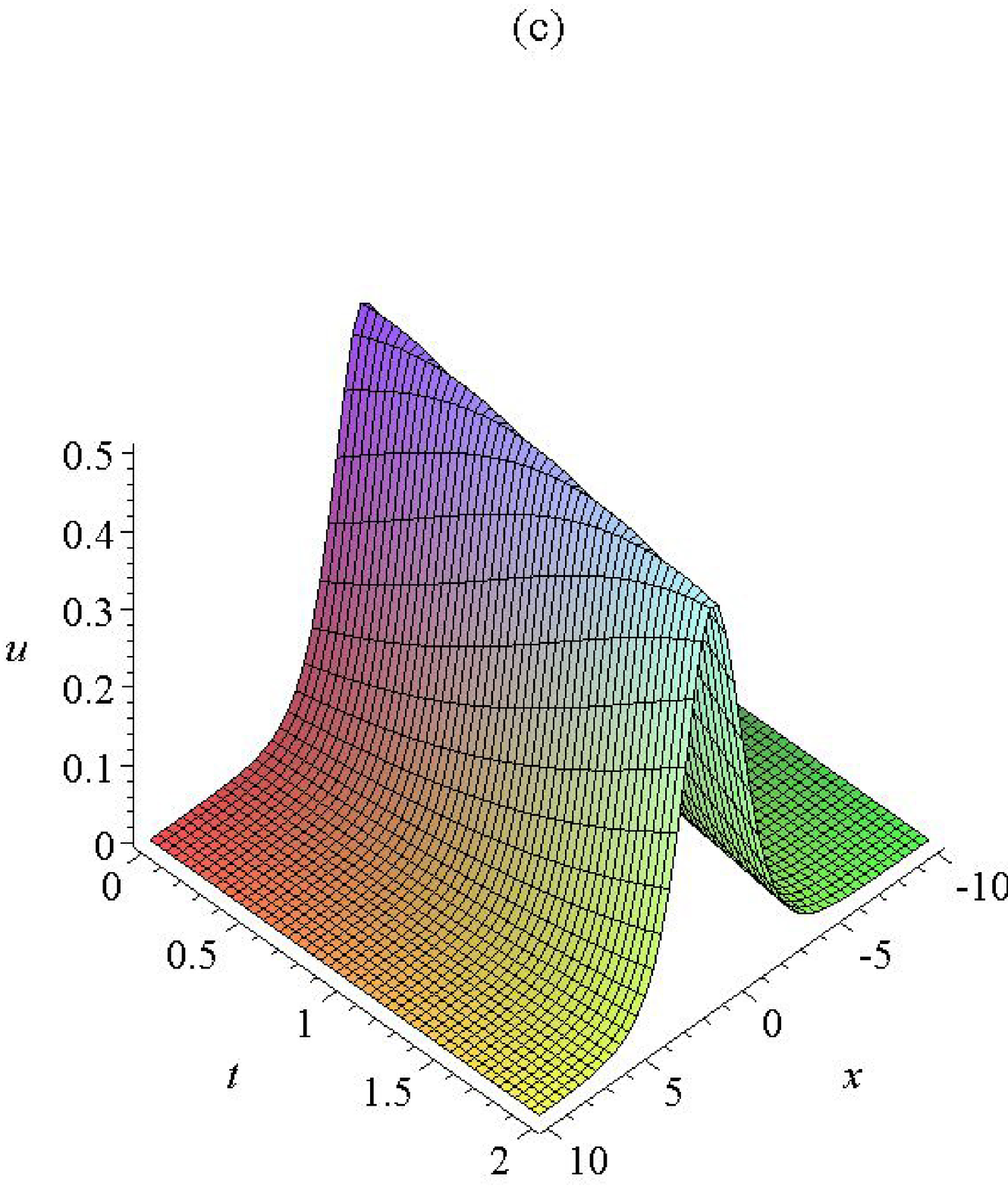}
\caption{Plots of the solitary wave like solutions.
{\bf (a)}\,solution $u_1^{(1)}$ with $p=-1,q=1,\omega=1.025$.
{\bf (b)} solution $u_2^{(2)}$ with $p=1,q=-1,\omega=0.995,\varepsilon=1$.
{\bf (c)} solution $u_1^{(4)}$ with $p=-1,q=1,\omega=1.025,\varepsilon=-1$.}
\label{fig:1}
\end{figure}
\par
(4)\,Taking (\ref{eq3:3}) with (\ref{eq2:3}),(\ref{eq2:11}) into (\ref{eq3:2})
and setting the coefficients of $F^iG^j~(i=1,2,3,4,5;j=0,1)$ to zero
we obtain the following set of algebraic equations
\begin{align*}
\left\{\begin{aligned}
&-{\frac {24a_2p^6r^2+96a_2p^4r^2+12b_2^2p^4r^2+48b_2^2p^2r^2}{q(p^2+2)^2}}+12a_2^2p=0,\\
&-{\frac {24a_2b_2p^4r^2+96a_2b_2p^2r^2}{q(p^2+2)^2}}
   +{\frac {24b_2p^9r^4+192b_2p^7r^4+384b_2p^5r^4}{q^2(p^2+2)^4}}=0,\\
&-{\frac {6a_1p^6r^2+24a_1p^4r^2+18b_1b_2p^4r^2+72b_1b_2p^2r^2}{q(p^2+2)^2}}\\
&{\quad}+30a_2p^2r+18b_2^2r+18a_2a_1p=0,\\
&-{\frac {6b_1^2p^4r^2+24b_1^2p^2r^2}{q(p^2+2)^2}}
-2{\omega}a_2p+6a_1^2p+6a_1p^2r-8a_2p^2q\\
&{\quad}+24b_2b1r-6b_2^2q+12a_0a_2p=0,\\
&-{\frac {18b_2a_1p^4r^2+72b_2a_1p^2r^2+18b_1a_2p^4r^2+72b_1a_2p^2r^2+60b_2p^5r^3
     +240b_2p^3r^3}{q(p^2+2)^2}}\\
&{\quad}+42a_2b_2r+{\frac {6b_1p^9r^4+48b_1p^7r^4+96b_1p^5r^4}{q^2(p^2+2)^4}}=0,\\
&{\frac {4b_1p^5r^2+16b_1p^3r^2}{(p^2+2)^2}}-15b_2prq+3b_1r^2p-3{\omega}b_2r-12a_2b_1q
     -12a_1b_2q+18a_1b_1r\\
&{\quad}+18a_0b_2r+{\frac {{\omega}b_1p^4r^2+4{\omega}b_1p^2r^2
     -6a_0b_1p^4r^2-24a_0b_1p^2r^2}{q(p^2+2)^2}}=0,\\
&{\frac {2{\omega}b_2p^4r^2+8{\omega}b_2p^2r^2-12b_1a_1p^4r^2-48b_1a_1p^2r^2-12a_0b_2p^4r^2
-48a_0b_2p^2r^2}{q(p^2+2)^2}}\\
&{\quad}-{\frac {12b_1p^5r^3+48b_1p^3r^3}{q(p^2+2)^2}}
    +{\frac {20b_2p^5r^2+80b_2p^3r^2}{(p^2+2)^2}}-18a_2b_2q+30a_2b_1r\\
&{\quad}+30a_1b_2r+30b_2pr^2=0,\\
&-a_1p^2q+6a_0a_1p-{\omega}a_1p+6b_1^2r-6b_1b_2q=0,\\
&-b_1pqr+b_2pq^2+6a_0b_1r-6a_0b_2q-6a_1b_1q-{\omega}b_1r+{\omega}b_2q=0.
\end{aligned}\right.
\end{align*}
Solution of this set of algebraic equations found by using Maple reads
\begin{align}
\label{eq3:11}
a_0={\frac {qp+\omega}{6}},a_1=-pr,
a_2={\frac {p^3r^2(p^2+4)}{q(p^2+2)^2}},b_1=0,
b_2={\frac {\varepsilon{rp^2}}{p^2+2}}\sqrt{-\frac {p(p^2+4)}{q}},
\end{align}
where $\varepsilon=\pm{1}$.\par
Taking (\ref{eq2:10}),(\ref{eq3:11}),(\ref{eq2:18}) into (\ref{eq3:3})
we obtain the Weierstrass type traveling wave solution of the KdV equation
\begin{align}
\label{eq3:12}
\begin{aligned}
u(x,t)&={\frac {144p\left(p^2q+{\omega}p+12q\right)\wp^2(\xi)+a\wp(\xi)
    +864q^2\varepsilon\sqrt{-\frac {p^3+4p}{q}}\wp^\prime(\xi)+a}{6\left(p^2q+12q+12p\wp(\xi)\right)^2}},\\
&a=24q\left(p^4q+{\omega}p^3-12p^2q+12{\omega}p-72q\right),\\
&b=q^2\left(p^5q+{\omega}p^4-36p^3q+24{\omega}p^2+144\omega\right),
\end{aligned}
\end{align}
where $\wp(\xi)=\wp(x-{\omega}t,g_2,g_3)$, the invariants $g_2$ and $g_3$ are
given by (\ref{eq2:12}).\par
Note that $pq < 0$ in (\ref{eq3:11}) and (\ref{eq3:12}), so (\ref{eq3:12}) can
only give the solitary wave solution of the KdV equation. Thus by substituting
the conversion formulas (\ref{eq2:13}) and (\ref{eq2:14}) with $\theta=-qp$
into (\ref{eq3:12}) we obtain the solitary wave like solutions of the KdV
equation as follows
\begin{align*}
\begin{aligned}
u_1^{(4)}(x,t)&={\frac {24p^2q\varepsilon\sqrt{p^2+4}\,\sinh\eta\cosh\eta
   +16(pq+\omega)\cosh^4\eta+r\cosh^2\eta+s}
   {6\left(4\cosh^2\eta+p^2\right)^2}},\\
u_2^{(4)}(x,t)&={\frac {-24p^2q\varepsilon\sqrt{p^2+4}\,\sinh\eta\cosh\eta
   +16(pq+\omega)\sinh^4\eta-r\sinh^2\eta+s}
   {6\left(4\sinh^2\eta-p^2\right)^2}},\\
&r=8p(-2p^2q+{\omega}p-6q),s=p^3(p^2q+{\omega}p+12q),
\end{aligned}
\end{align*}
where $\eta={\frac 12}\sqrt{-pq}(x-{\omega}t),qp<0$.\par
\begin{figure}[h]
\centering
\includegraphics[width=0.30\textwidth]{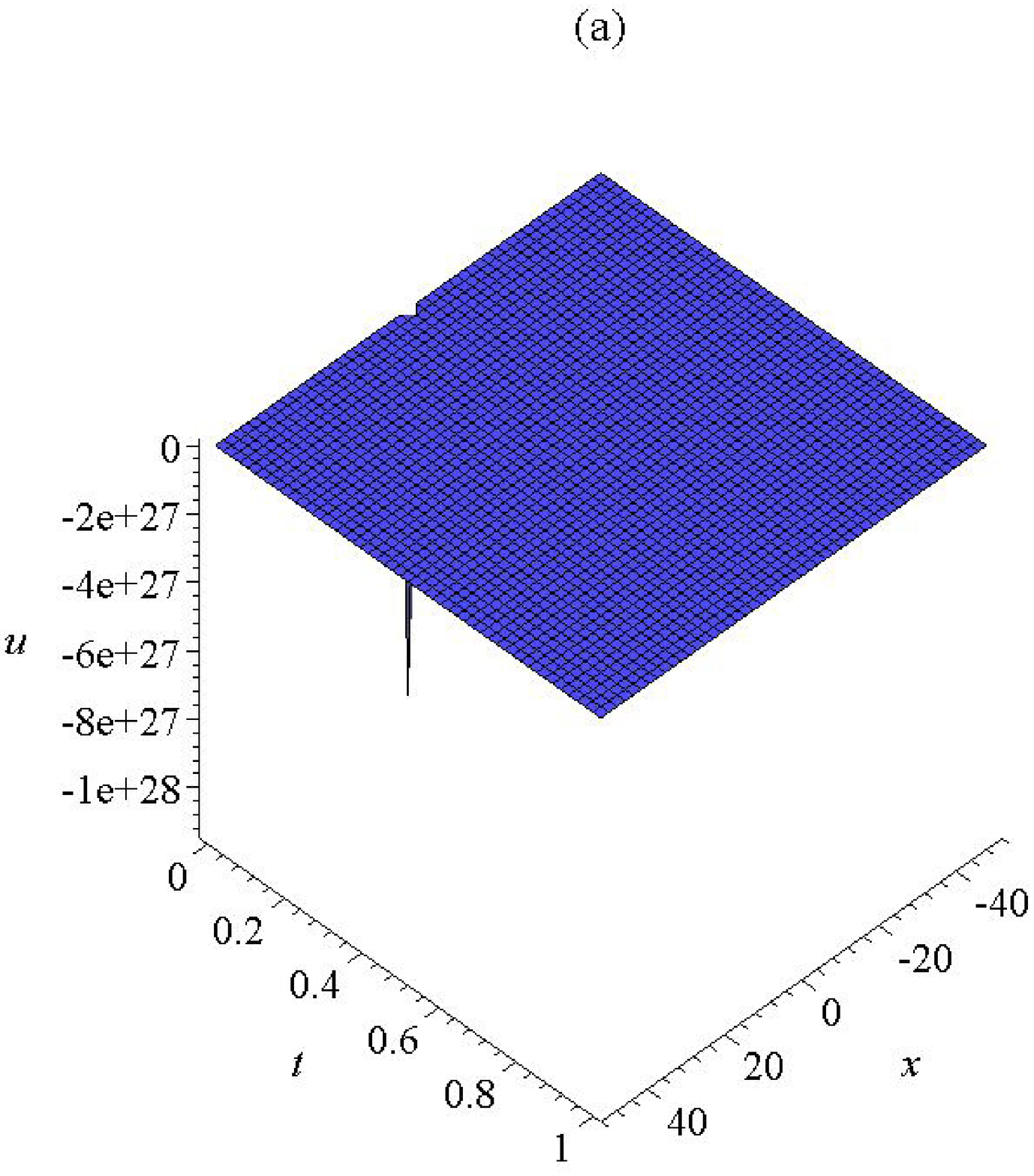}
\includegraphics[width=0.30\textwidth]{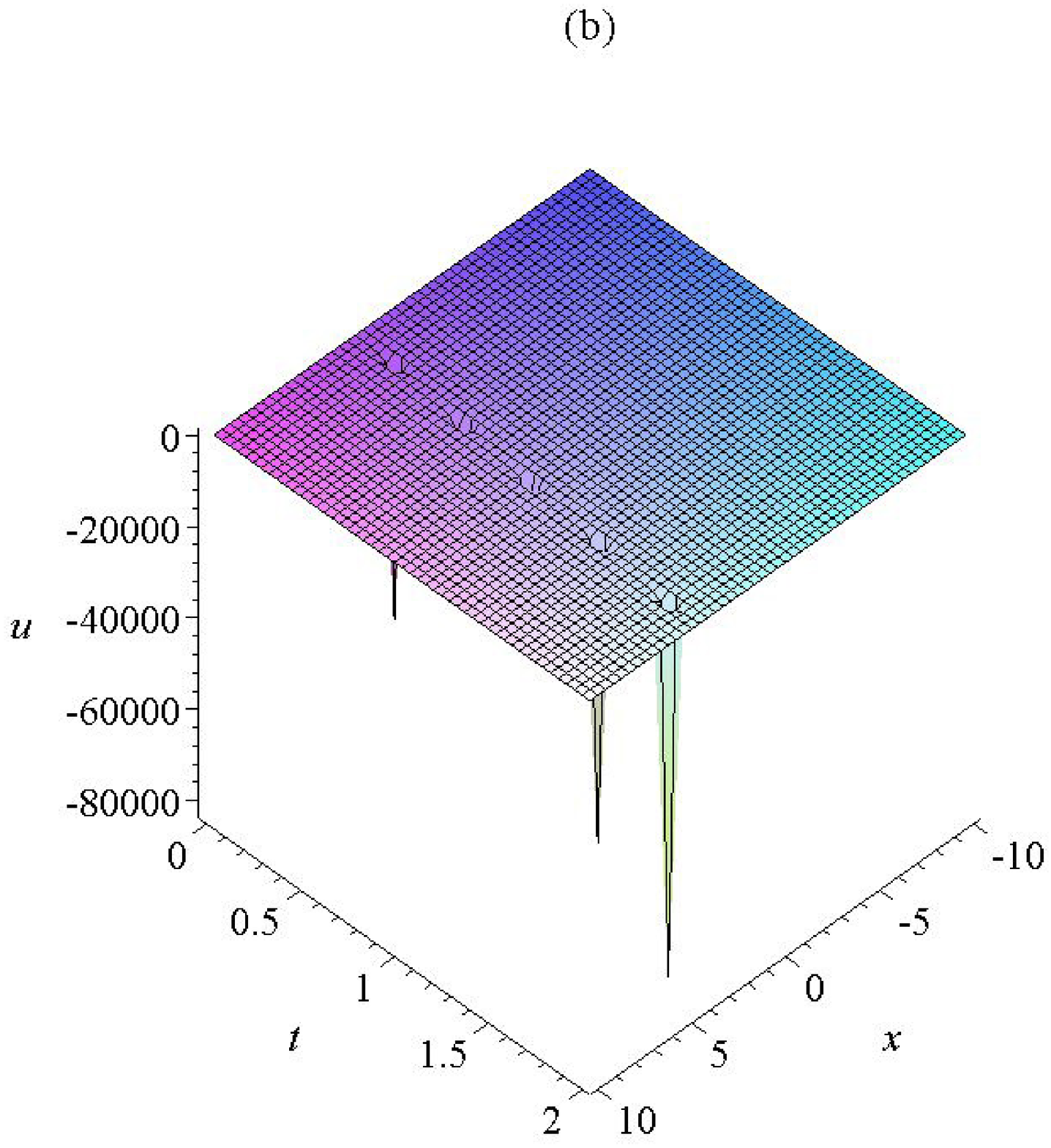}
\includegraphics[width=0.30\textwidth]{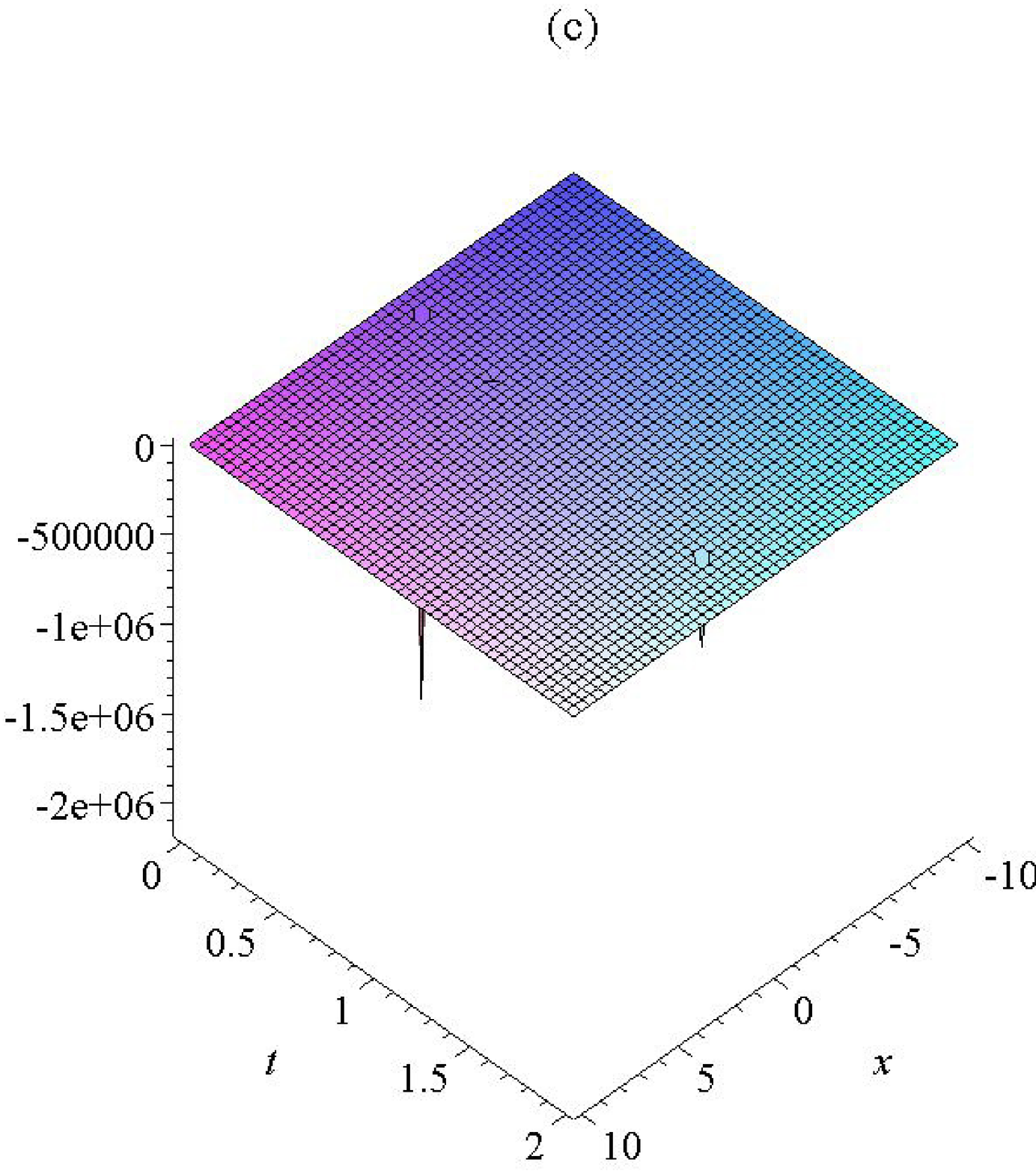}
\caption{Plots of the singular solitary wave solutions.
{\bf (a)}\,solution $u_2^{(1)}$ with $p=1,q=-1,\omega=1.025$.
{\bf (b)} solution $u_1^{(2)}$ with $p=-1,q=1,\omega=1.025,\varepsilon=-1$.
{\bf (c)} solution $u_2^{(4)}$ with $p=1,q=-1,\omega=1.025,\varepsilon=1$.}
\label{fig:2}
\end{figure}
\begin{figure}[h]
\centering
\includegraphics[width=0.45\textwidth]{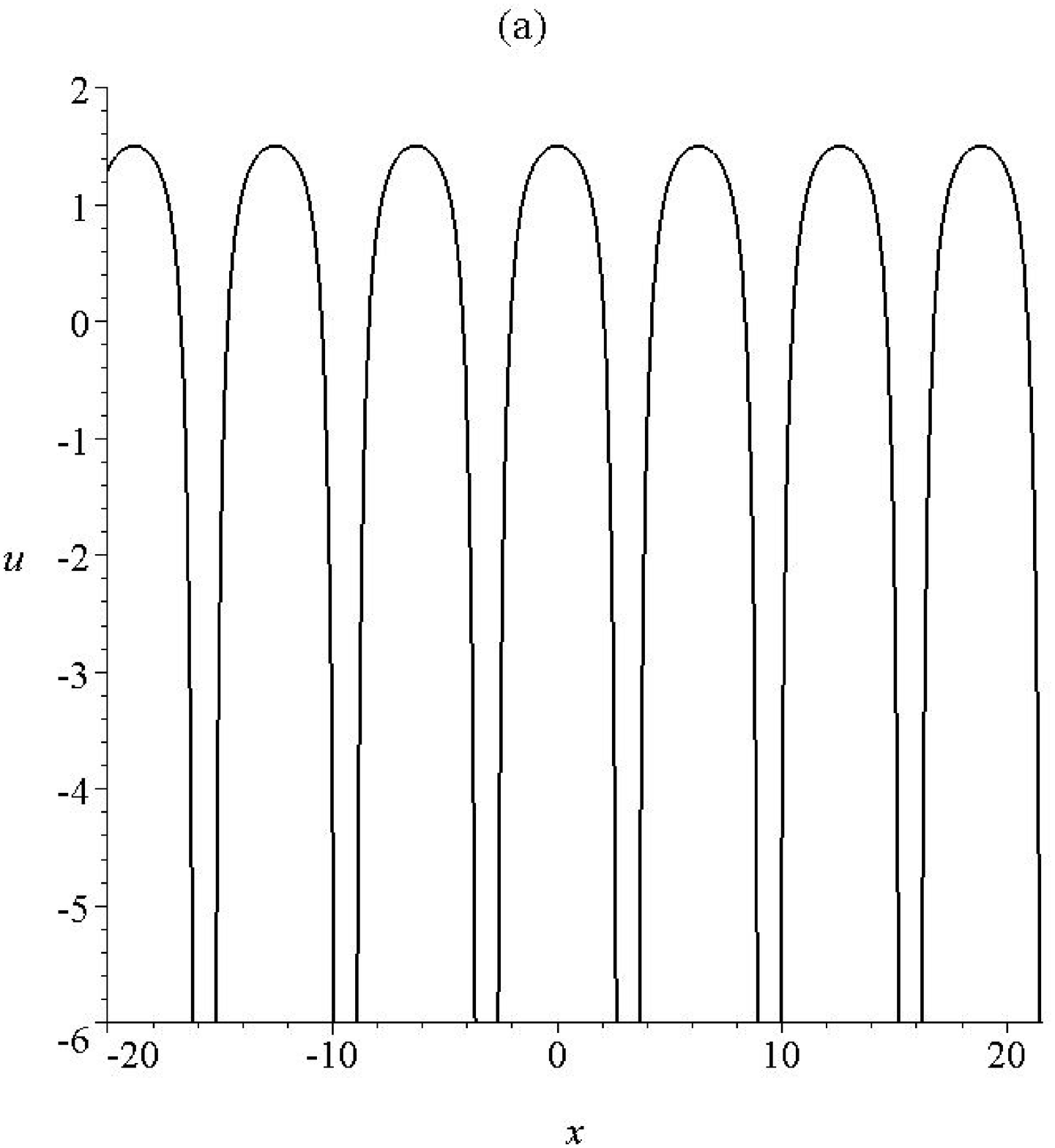}
\includegraphics[width=0.45\textwidth]{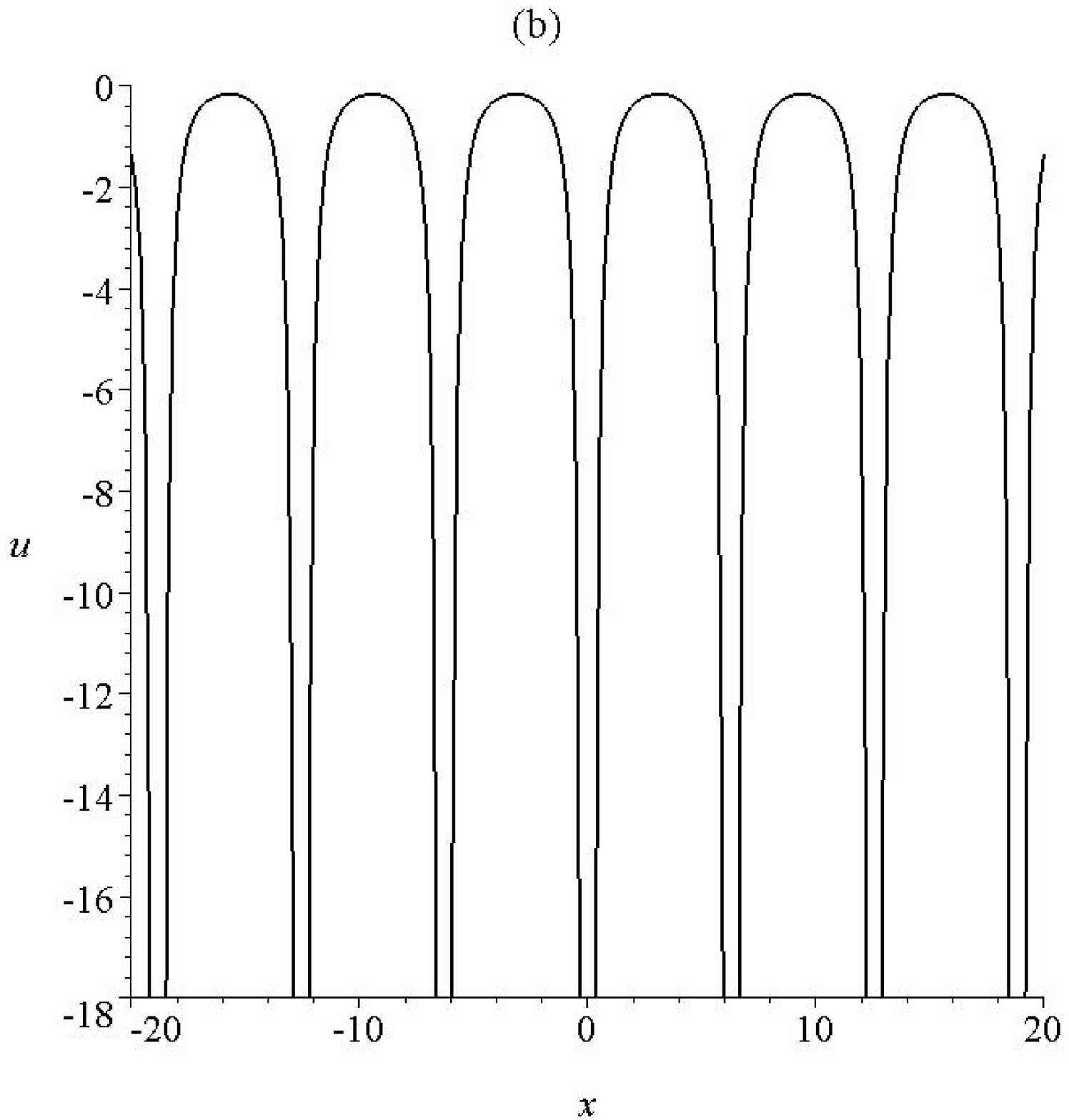}
\caption{Plots of the periodic solutions.
{\bf (a)}\,solution $u_3^{(1)}$ with $p=1,q=1,\omega=1.025$.
{\bf (b)} solution $u_4^{(1)}$ with $p=1,q=1,\omega=1.025$.}
\label{fig:3}
\end{figure}

It is obvious that the obtained solutions of the KdV equation are
related to the coefficients $p,q$ and $r$ of the projective Riccati
equations (\ref{eq2:3}). The solitary wave like solutions,
the singular solitary wave solutions and the periodic solutions
of the KdV equation are plotted in Fig.\ref{fig:1},Fig.\ref{fig:2}
and Fig.\ref{fig:3}.The graphs of the solutions $u_i^{(3)}\,(i=1,2,3,4)$
are completely similar to that of the solutions $u_i^{(1)}\,(i=1,2,3,4)$,
so it is omitted.
\section{Conclusions}
\label{sec:4}
In the present paper, the Weierstrass type projective Riccati equation
expansion method is proposed to construct Weierstrass elliptic function
solutions of NLEEs. At the same time, the conversion formulas are also
proposed to transform these Weierstrass elliptic function solutions into
the hyperbolic and trigonometric function solutions of NLEEs.
Our method is more applicable than the projective Riccati equation
expansion method and it can be regarded as an extension of the projective
Riccati equation expansion method. In order to explain our method more clear
we need to point out the following three points.\par
(1)\,The Weierstrass elliptic function solutions (\ref{eq2:4}) and (\ref{eq2:6})
of the projective Riccati equations are known~\cite{RefJ-8a,RefJ-8b,RefJ-8c},
and the solutions (\ref{eq2:8}) and (\ref{eq2:10}) are new.\par
(2)\,The known conversion formula such as~\cite{RefJ-8a,RefJ-8b}
\begin{align}
\label{eq4:1}
\wp(\xi,g_2,g_3)=e_2-(e_2-e_3)\cn^2\left(\sqrt{e_1-e_3}\,\xi,m\right),
\end{align}
is related to the roots $e_1,e_2,e_3$ of the third order polynomial
equation $w^3-g_2w-g_3=0$,but our conversion formulas (\ref{eq2:13})--
(\ref{eq2:16}) don't require these roots. Using formula (\ref{eq4:1}),
we can transform the Weierstrass elliptic function solutions of NLEEs
into the Jacobian elliptic function solutions, and then the hyperbolic
and trigonometric function solutions are obtained by taking the limit of
modulus $m\rightarrow{1}$ and $m\rightarrow{0}$, but they may be
incorrect~\cite{RefB-1,RefB-2}.
In contrast, our formulas (\ref{eq2:13})--(\ref{eq2:16}) can directly convert
the Weierstrass elliptic function solutions of NLEEs into the hyperbolic
and trigonometric function solutions, and ensure that the obtained solutions
are correct.
In addition, the conversion formulas (\ref{eq2:13})--(\ref{eq2:16}) also
can be used in other Weierstrass elliptic function methods to transform
the Weierstrass elliptic function solutions into the hyperbolic and trigonometric
function solutions.\par
(3)\,Although we have concerned with the KdV equation, our method can be
applied to construct the exact solitary wave and periodic wave solutions
of a wide class of NLEEs.

\end{document}